\documentclass[preprint,prb,final]{revtex4}

\usepackage[T1]{fontenc}
\usepackage{graphicx}
\usepackage{amsfonts}
\usepackage{amssymb}
\usepackage{amsmath}
\usepackage{color}

\begin{document}

\title{Non-reciprocal Oersted field contribution to the current-induced frequency shift of magnetostatic surface waves}

\author{Mohammad Haidar}
\email{mohammad.haidar@hotmail.fr}
\altaffiliation[Present address: ]{Physics Department, University of Gothenburg, 412 96, Gothenburg, Sweden}
\author{Matthieu Bailleul}
\email{matthieu.bailleul@ipcms.unistra.fr}
\affiliation{Institut de Physique et Chimie des Mat\'{e}riaux de Strasbourg - IPCMS and NIE, UMR 7504 CNRS - Universit\'{e} de Strasbourg, 23 rue du Loess, BP 43, 67034 Strasbourg Cedex 2, France}
\author{Mikhail Kostylev}
\affiliation{School of Physics, The University of Western Australia, Crawley, WA 6009, Australia}
\author{Yuyan Lao}
\affiliation{University of Science and Technology, Hefei, China \\
and \\
School of Physics, The University of Western Australia, Crawley, WA 6009, Australia}

\date{\today}

\begin{abstract}
The influence of an electrical current on the propagation of magnetostatic surface waves is investigated in a relatively thick (40 nm) permalloy film both experimentally and theoretically. Contrary to previously studied thinner films where the dominating effect is the current-induced spin-wave Doppler shift, the magnetic field generated by the current (Oersted field) is found to induce a strong non-reciprocal frequency shift which overcompensates the Doppler shift. The measured current induced frequency shift is in agreement with the developed theory. The theory relates the sign of of the frequency shift to the spin wave modal profiles. The good agreement between the experiment and the theory confirms a recent prediction of a counter-intuitive mode localization for magnetostatic surface waves in the dipole-exchange regime.
\end{abstract}

\maketitle

\section{INTRODUCTION}
The spin waves are the elementary magnetic excitations of ferromagnets. Although they are known for a long time, their study at the nanometer scale in thin films is the subject of a recent field of research called magnonics,\cite{Kruglyak2010} which proposes to use them as information vectors for future applications in data storage and signal processing.\cite{Hertel2004,Kostylev2005,Khitun2008} Another field of research where spin waves play an important role is that of spin transfer torque, through the phenomenon of the current-induced spin-wave Doppler shift (CISWDS): when an electrical current flows along a metal ferromagnet in which a spin wave is excited, there is a transfer of angular momentum along the spin wave propagation direction, which shifts the spin-wave frequency by an amount proportional to the degree of spin-polarization of the current.\cite{Vlaminck2008} The CISWDS can therefore be used to probe directly spin-polarized electron transport in various experimental conditions and materials.\cite{Zhu2010,Zhu2011,Thomas2011,haidar2013} It was also suggested that another spin-torque effect (the current-induced modification of the spin-wave attenuation) could be used to amplify them.\cite{Seo2009,Sekiguchi2012} It is essential to understand precisely the influence of the electrical current onto the propagation of the spin wave in order to be able to rule out possible concurrent physical effects which are likely to combine with the spin transfer torque, in particular the effect of the inhomogeneous magnetic field generated directly by the electrical current (the Oersted field). For future development in these two fields (magnonics and spin-wave spin-transfer torque) a good understanding of the fundamental physics of spin wave propagation in metallic ferromagnetic films and of the influence of a DC electrical current on it is therefore needed.

The most relevant configuration for experimental studies of spin wave propagation is the so-called MagnetoStatic Surface Wave (MSSW) configuration (also known as Damon-Eschbach configuration) in which the equilibrium magnetization $\textbf{M}$ and the spin-wave wave vector $\textbf{k}$ are perpendicular to each other, and both lie in the plane of the film.\cite{Stancil2009,Damon1961} This configuration has two advantages: (i) because $\textbf{M}$ is oriented in the film  plane, moderate magnetic fields are sufficient to magnetize the film, (ii) because $\textbf{M}$ is perpendicular to $\textbf{k}$, and lies in the film plane, the precession of magnetization induces two components of the dynamic demagnetizing field: in the film plane and perpendicular to it, both with a strong dependence on $\mid{\bf{k}}\mid$. This unique structure of the dynamic demagnetizing field translates into a relatively high group velocity. Due to the large group velocity, for a given relaxation time, these waves propagate quite far before they completely die off due to the attenuation in the medium. This significantly facilitates the measurements with respect to the other spin wave configurations. However, MSSW also has a very specific property called non-reciprocity: the amplitudes, mode profiles and frequencies of the waves travelling in the two opposite propagation directions ($k>0$ and $k<0$) do not coincide. The amplitude non-reciprocity is a property related to MSSW excitation by external energy sources: the efficiency of excitation of spin waves by a microstrip or coplanar inductive antenna located on the film surfaces is larger for one propagation direction than for the opposite one.\cite{Schneider2008} The modal profile non-reciprocity manifests itself in the fact that these spin waves have a larger amplitude on one side of the film than on the other one (surface character of the wave). The surface at which the wave is localized swaps upon reversal of the propagation direction.\cite{Stancil2009} Finally, frequency non-reciprocity may also be present whenever the film is asymmetric in the thickness direction.\cite{Amiri2007} For a long time the non-reciprocity of MSSW has been studied in thick, low magnetization Yttrium Iron Garnet (YIG) films.\cite{Stancil2009,Schneider2008} The investigations of the MSSW non-reciprocity for the thin, high magnetization permalloy (Py) films used in most magnonics studies\cite{bailleul2003,Amiri2007,Demidov2009,Sekiguchi2010} are more recent. Quite recently it has been shown theoretically that for a given applied-field direction, MSSW in thin Py films may be localized at the film surface opposite to the one of MSSW localisation in thick YIG films due to the more pronounced role of the exchange interaction in the magnetization dynamics in the Py films.\cite{Kostylev2013}

In the context of the studies of the current-induced spin wave Doppler shift, the amplitude and frequency MSSW non-reciprocities complicate the extraction of the Doppler shift, because signals corresponding to counter-propagating spin-waves cannot be directly compared contrary to the first CISWDS measurement which dealt with reciprocal spin waves.\cite{Vlaminck2008} Different procedures have been proposed to extract the Doppler shift, either by combining measurements taken at different polarities of $M$ and $k$,\cite{Zhu2010} or by combining measurements taken at different polarities of $I$ and $k$.\cite{haidar2013}

In this paper, we build upon these previous works. We measure very precisely the non-reciprocity of propagation of magnetostatic surface waves and its modification by an electrical current in a permalloy film which is thicker (40 nm) than the ones employed in the previous studies (6 nm-20 nm in Refs. \onlinecite{Zhu2010,Sekiguchi2012,haidar2013}). Surprisingly, we observe that the current-induced frequency shift behaves very differently from what is observed for the thinner films: it does not scale linearly with the wave-vector and can even change its sign. We attribute this behavior to a large non-reciprocal contribution from the Oersted field of the DC current which combines with the Doppler effect. To observe a noticeable Doppler frequency shift, large densities of DC current are required. For the same large current density the total current through a thicker film is larger which results in a larger Oersted field than for a thinner film. Furthermore, for the same wave number the mode profile asymmetry is also larger for a thicker film, so that the contribution of the Oersted field to the mode frequency becomes non-negligible with respect to the Doppler frequency shift. Our study also reveals that the Oersted-field and Doppler contributions have the opposite signs. It is impossible to distinguish between the two contributions by employing symmetry considerations because they behave in the same way as functions of the directions of $\textbf{I}$, $\textbf{k}$ and $\textbf{H}$. On the other hand, as we show here, they have different dependencies on the magnitude of the wave vector. The total frequency shifts calculated using the modal profiles described in Ref.\onlinecite{Kostylev2013} are in good agreement with the measured ones. Because the non-reciprocal Oersted contribution to the current-induced frequency shift is very sensitive to the mode profile asymmetry, our observation provides a confirmation of the counter-intuitive MSSW localization behavior predicted in Ref.\onlinecite{Kostylev2013}.

The paper is organized as follows. The experimental results are presented in section \ref{sec:exp}. In section \ref{sec:qualitat}, we provide a qualitative interpretation of the measured current-induced frequency shifts. The theoretical calculations are presented in section \ref{sec:theory} and we conclude in section \ref{sec:conclusion}.

\section{EXPERIMENTAL RESULTS}\label{sec:exp}

\subsection{Propagating spin wave measurements for $I=0$}\label{subsec:psws}

\begin{figure}[!htp]
\begin{center}
\includegraphics[scale=0.4]{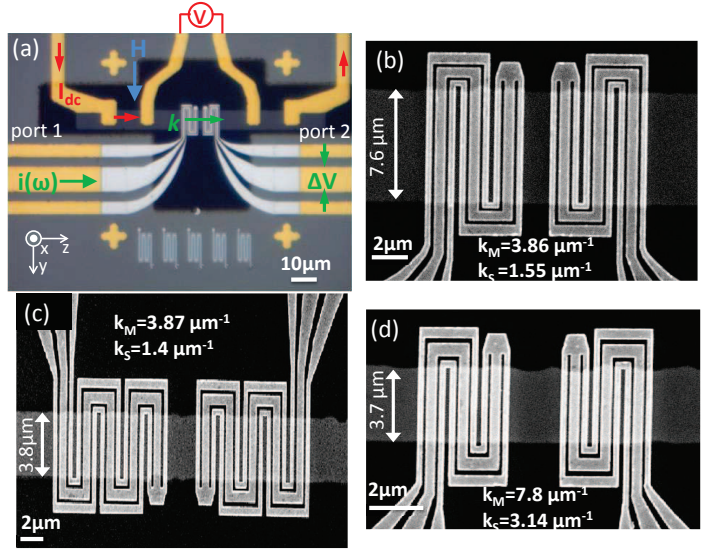}
\caption{(a) Optical microscope image of one CISWDS device. One recognizes the Permalloy strip ion-milled from a continuous film ($Al_2O_3$ 21 nm / Py 40nm / $Al_2O_3$ 5nm), the four DC current pads and the two coplanar waveguides (Ti 10 nm / Au 60 nm), the insulating spacer (SiOx 120 nm) and the two spin-wave antennae (Ti 10 nm / Al 120 nm). The conventions used in the text for the directions of positive $\bf{k}$, $\bf{I}$ and $\bf{H}$ are shown. (b),(c),(d) Scanning electron microscope images showing the strip and the antennae for each of the three fabricated devices.}\label{fig:1}
\end{center}
\end{figure}
The sample used in the experiment consists of a 40 nm-thick permalloy (Py) film sandwiched between $Al_{2}O_{3}$ layers. It was grown on an intrinsic silicon substrate by magnetron sputtering. The chip contains several devices of the type shown in Fig. \ref{fig:1}(a). Each device comprises a Py strip of width $w$ and a pair of narrow-band microwave spin wave antennae of meander shape for the excitation and detection of spin waves with wave vector $k$. The antennae are separated from the strip by a 120 nm thick SiOx insulating layer. In addition, four DC pads are connected to the strip in order to launch a DC current $I$ into it and to measure its resistance. Figs. \ref{fig:1}(b-d) show scanning electron microscopy images of each device. In each panel, we indicate the strip width and the characteristic wave vector for each device. Spin waves are excited by the antenna with a main excitation peak centered at a wave vector $k_{M}$ and a secondary peak centered at a lower wave vector $k_{S}$ as described in the appendix of Ref. \onlinecite{Vlaminck2010}. The samples are placed in a uniform static magnetic field $\mathbf{H}$ applied in the film plane, perpendicular to the propagation direction of the spin waves, which corresponds to the magnetostatic surface wave configuration. The propagating spin wave spectroscopy (PSWS) measurements are performed as described in detail elsewhere.\cite{haidar2013,haidarPhD,Vlaminck2010}

\begin{figure}[!htp]
\begin{center}
\includegraphics[scale=0.5]{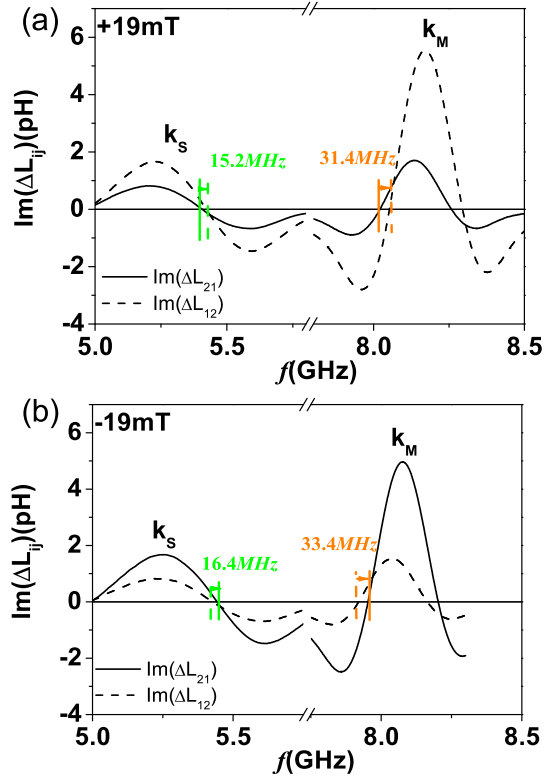}
\caption{The imaginary parts of the mutual inductance signals for $k>0$ and $k<0$ propagations (solid and dashed curves respectively). The signals were measured for the device shown in Fig. \ref{fig:1}(c). The strip width is $w= 3.8$ $\mu m$ and the characteristic wave numbers are $k_{M}=3.86$ $\mu m^{-1}$ and $k_{S}=1.4$ $\mu m^{-1}$. The measurements were performed in an external field (a) $H=19$ mT and (b) $H=-19$ mT in the absence of the DC current.}\label{fig:2}
\end{center}
\end{figure}

The devices were first characterized in the absence of the DC current. Fig. \ref{fig:2}(a) shows typical mutual inductance signals: $\Delta L_{21}$ (solid curve) which corresponds to a wave propagating from port 1 to port 2 [$k>0$, see notations in Fig. \ref{fig:1}(a)] and $\Delta L_{12}$ (dashed curve) which corresponds to a wave propagating from port 2 to port 1 ($k<0$). For both signals, we observe two distinct wave packets centered at $5.4$ GHz and $8.1$ GHz. These frequencies are in good agreement with the values expected from the MSSW dispersion relation with wave vectors $k_{M}$ and $k_{S}$ respectively.\cite{Stancil2009} One also notices that the transmitted amplitude for $k<0$ is higher than one for $k>0$ for both peaks. The ratio of the amplitude of the $k<0$ signal to the $k>0$ signal is about 3 and 2 for the $k_{M}$ and $k_{S}$ peaks respectively. This amplitude asymmetry is in agreement with Refs. \onlinecite{Schneider2008,Demidov2009,Sekiguchi2010}, where it was explained based on differences in elliptical polarizations of the oscillating magnetization of the spin wave and of the microwave field generated by the antennae. The spin wave whose magnetization precession has the same polarization as the driving microwave field is excited more strongly than the spin wave with the opposite polarization. In agreement with the theoretical expectation,\cite{emtage1978} the wave with higher amplitude is propagating with a wave vector $\mathbf{k}\parallel(\mathbf{n}\times \mathbf{M})$ where $\mathbf{n}$ is the internal normal to the film surface close to which the antenna is located, which corresponds in our experiment to the $k<0$ signal for the $+H$ field orientation. We also observe that the $k<0$ signal lies at a slightly higher frequency than the $k>0$ signal. The frequency shift is about $31.4$ MHz and $15.2$ MHz for $k_{M}$ and $k_{S}$ peaks respectively. As in the case of thinner films,\cite{haidar2013} we attribute this frequency non-reciprocity to the combination of the modal profile non-reciprocity with some asymmetry of the magnetic properties of the films with respect to its mid-plane (e.g. a different surface anisotropy at the top and bottom interfaces\cite{hillebrands1990} or an inhomogeneous magnetization distribution across the film thickness\cite{Vohl1989,Kostylev2010}). A quantitative interpretation of this feature is left for future work because it would require a very accurate knowledge of the film structure. As we switch the direction of the static field to the negative one, we observe that the $k>0$ and $k<0$ signals swap their amplitudes and frequencies, which means that the $k>0$ signal now has a higher amplitude and a higher frequency with respect to the $k<0$ signal, as shown in Fig. \ref{fig:2}(b). Hence, the amplitude and frequency non-reciprocities reverse when the direction of the external field is reversed. This is in good agreement with the interpretation given above because both the polarization of the oscillating magnetization and the modal profile asymmetry are expected to reverse when the equilibrium magnetization is switched but not the polarization of the microwave field of the antennae.

\subsection{Current-induced modifications of the spin-wave signals}\label{subsec:CIFS}

\begin{figure}[!htp]
\begin{center}
\includegraphics[scale=0.6]{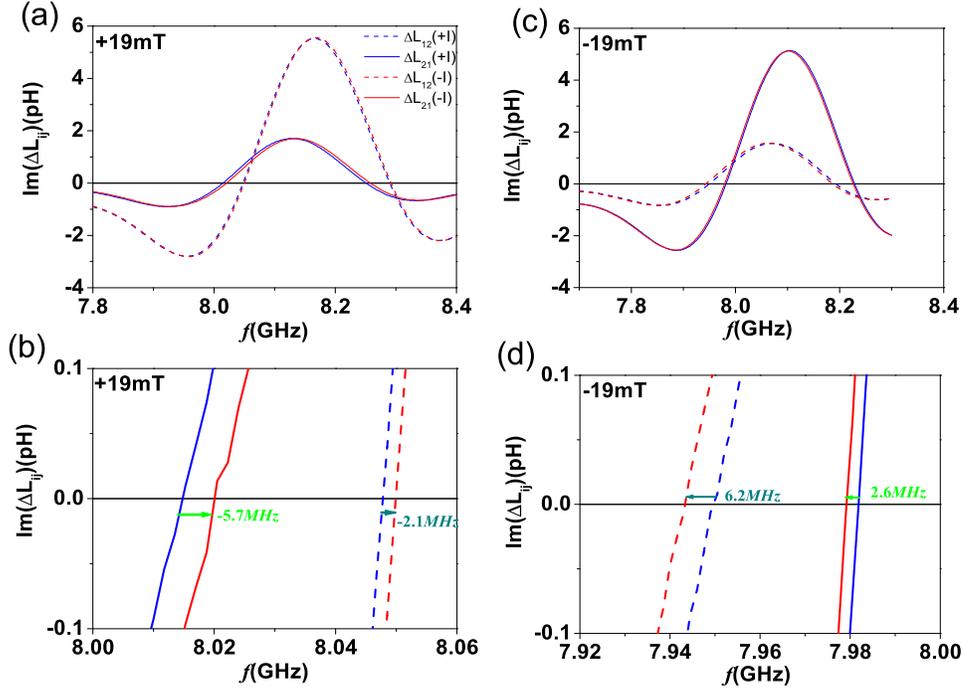}
\caption{The mutual inductance signals measured in the presence of the DC current. The measurements were performed at (a) $+19$ mT and (c) $-19$ mT at $|I|=7.5$ mA for the same device as Fig. \ref{fig:2}. (b),(d) Zoomed views of the signals showing the current induced frequency shifts.}\label{fig:3}
\end{center}
\end{figure}

\begin{figure}[!htp]
\begin{center}
\includegraphics[scale=0.5]{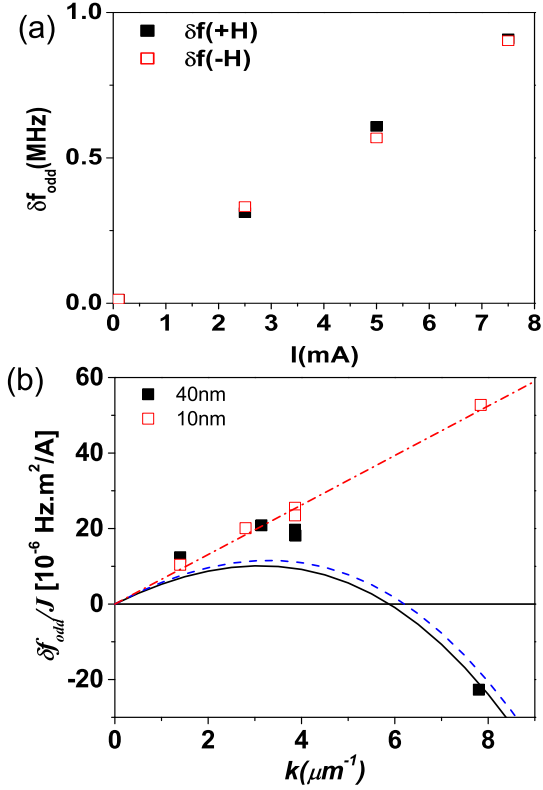}
\caption{(a) Variation of $\delta f_{odd}$ as a function of the electrical current (same device as in Figs. \ref{fig:2},\ref{fig:3}). (b) Ratio of $\delta f_{odd}$ over the current density $J$ as a function of the wave vector for 40 nm (filled squares) and 10nm (open squares) thin films. Lines: respective theoretical data. Solid black line: rigorous numerical calculation in real space for the 40 nm-thick film. Dashed blue line: the same, but analytical solution using Eq. (\ref{eq:19}). Dash-dotted red line: film thickness is 10 nm (for this thickness both results of the rigorous numerical calculation and of the analytical formula coincide to graphical accuracy with a linear dependence corresponding to the CISWDS only).}\label{fig:4}
\end{center}
\end{figure}

Let us now investigate the effect of an electrical current on the propagating spin waves. Figure \ref{fig:3}(a) shows the mutual-inductance spectra recorded in the presence of an electrical current $I=\pm7.5$ mA. The small current-induced frequency shifts are better seen in Fig. \ref{fig:3}(b) which shows a zoom close to the intercept of the curves with the horizontal axis. The $+I$ curves (blue lines) appear to be at slightly smaller frequencies than the $-I$ curves (red lines) and the shift is clearly higher for $\Delta L_{21}$ (solid curve). As in Ref. \onlinecite{haidar2013}, we define $\delta f_{ij}=f_{ij}(+I)-f_{ij}(-I)$ where $f_{ij}(I)$ is the frequency at which the $Im \Delta L_{ij}(I)$ signal ($i,j=1,2$) vanishes. One obtains $\delta f_{12}=-2.1$ MHz and $\delta f_{21}=-5.7$ MHz. These two values are combined as follows: $\delta f_{even}=(\delta f_{12}+\delta f_{21})/4=-1.95$ MHz is the part of the current-induced frequency shift which is even in $k$ and $\delta f_{odd}=(\delta f_{12}-\delta f_{21})/4=+0.9$ MHz is the part of the current-induced frequency shift which is odd in $k$. Fig. \ref{fig:3}(c,d) show the PSWS signals measured when the direction of the magnetic field is switched. The current induced frequency shifts are now $\delta f_{21}=+2.6$ MHz and $\delta f_{12}=+6.2$ MHz which gives $\delta f_{odd}=+0.9$ MHz and $\delta f_{even}= +2.2$ MHz. Apparently, the part of the current induced frequency shift that is even in $k$ is also odd in $H$, and the part which is odd in $k$ is even in $H$. The current induced frequency shifts also scale linearly with the DC current. This is exemplified in Fig. \ref{fig:4}(a) for $\delta f_{odd}$.\\

Before we discuss further $\delta f_{odd}$, which is the part that contains the CISWDS (the Doppler shift changes sign between two counter-propagating spin-waves) and also the non-reciprocal Oersted field contribution we will discuss below, let us discuss briefly $\delta f_{even}$. As for thinner films,\cite{haidar2013} we attribute it to a (reciprocal) Oersted field contribution induced by a top/bottom asymmetry of the ferromagnetic metal film: If the electrical properties are not perfectly symmetric with respect to the film midplane (e.g. the top part is slightly more conductive than the bottom part), the Oersted field is not entirely antisymmetric with respect to the film midplane and does not average out to zero, so that a small residual field will add to or subtract from the applied magnetic field and therefore modify the frequency. Similarly, if the magnetic properties of the film are not perfectly symmetric (e.g. magnetization pinning is stronger at the top than at the bottom surface), there might be a slight (reciprocal) asymmetry of the spin-wave profile with respect to the sample mid-plane, so that a perfectly antisymmetric Oersted field weighted by this profile would not average out to zero. In this picture, the asymmetry occurring across the film thickness originates from the film itself, it is therefore not expected to reverse when $k$ is reversed. These effects are thus expected to be reciprocal and to lead to a current-induced frequency shift even in $k$. Because the Oersted field combines vectorially with the external field $H$, this contribution is also expected to be odd in $H$, as it was deduced from Fig. \ref{fig:3}(b) and (d). A quantitative understanding of this even contribution is beyond the scope of this paper because it would require a very detailed knowledge of the distribution of the material properties over the film thickness.

Let us now focus on the part of the current-induced frequency shift which is odd in $k$. The current-induced spin-wave Doppler shift $\Delta f_{Dop}$ expected to contribute to this part writes:\cite{Vlaminck2008}
\begin{equation}\label{eq:1}
\Delta f_{Dop}=-\frac{\mu_{B}}{2\pi |e|} \frac{P}{M_{s}} \frac{I}{L w} k,
\end{equation}
where $w$ and $L$ are the width and the thickness of the ferromagnetic metal strip, $\mu_{B}$ is the Bohr magnetron, $|e|$ is the magnitude of the electron charge, $M_{s}$ is the saturation magnetization and $P=\frac{J_{\uparrow}-J_{\downarrow}}{J_{\uparrow}+J_{\downarrow}}$ is the degree of the spin-polarization of the electrical current. To explore the wave vector dependence of $\delta f_{odd}$, we compare the current-induced frequency shifts measured on the three devices (main excitation peaks at $k_M=3.9$ and $7.8$ $\mu m^{-1}$) and we use also the current-induced shifts measured for the secondary peaks ($k_S=1.4$ to $3.14$ $\mu m^{-1}$). To account for the different in the strip widths of the devices, $\delta f_{odd}$ is actually plotted as a function of the current density $J=I/Lw$. The slopes of the linear fits obtained in each case are plotted in Fig. \ref{fig:4}(b) as a function of the wave vector, together with the data points obtained following the same procedure for a $10$ nm film. The difference between the two film thicknesses is obvious. For the 10 nm film, $\delta f_{odd}$ scales linearly with the wave vector, as expected from the Doppler effect [see Eq. (\ref{eq:1})]. On the other hand, for the 40 nm film, $\delta f_{odd}$ first increases between $k=1.4$ and $3.4$ $\mu m^{-1}$, then saturates and finally decreases strongly to become negative at $k=7.8$ $\mu m^{-1}$. A direct application of Eq. (\ref{eq:1}) leads to the following evaluates for the degree of spin-polarization $P$: $0.52\pm0.02$ for $L=10$ nm, a value which can be understood by considering the spin-polarized electron scattering processes acting in a permalloy thin film,\cite{haidar2013} and between $0.73$ and $-0.24$ for $L=40$ nm (the two values corresponding to $k=1.4$ and $7.8$ $\mu m^{-1}$ respectively). We believe this latter range of values does not make sense: there is no reason for the degree of spin-polarization to depend on the wave vector in this range (the spin-wave wavelength $\lambda=0.8-4.5$ $\mu m$ remains much larger than any of the characteristic lengths for electrical transport in such a film) and there is no reason for it to become negative (the dominant electron scattering processes at large thickness is the scattering by the alloy disorder, which is known to give rise to a strong positive spin-polarization\cite{Banhart1997}). So we believe that another effect combines with the CISWDS to generate the $\delta f_{odd}$ we measure.

\section{QUALITATIVE EXPLANATION}\label{sec:qualitat}

In this section we present a naive qualitative picture which explains the experimentally observed nonlinear wave-number dependence of the current-induced frequency shift which translates into the unphysical wave-number dependence of the extracted degree of spin polarization together with its unphysical change of sign.  We claim that the wave number dependence of the frequency shift is due to one more process taking place in parallel to the Doppler effect. This is the "Oersted-field induced frequency shift" (OFIFS). It was previously mentioned in Ref. \onlinecite{Sekiguchi2012}, but has not been explored in detail yet. The idea is that the Oersted field generated by the DC current $H_{Oe}$ can modify the spin wave dispersion in a metallic ferromagnetic film and make the spin wave dispersion non-reciprocal.

\begin{figure}[!htp]
\begin{center}
\includegraphics[scale=0.50]{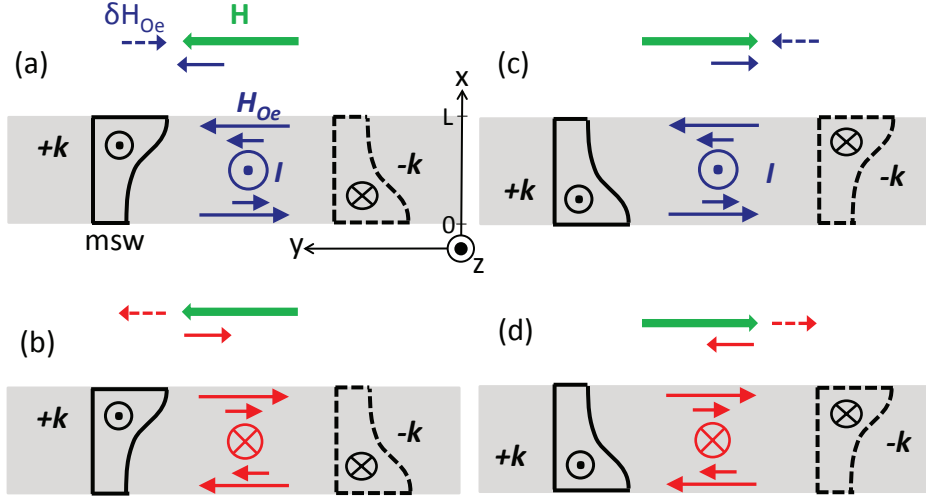}
\caption{A qualitative sketch of the non-reciprocal Oersted field induced frequency shift. It shows the spin-wave modal profile for $k>0$ (black full lines) and $k<0$ (black dashed lines) under various orientation of $H$ (in green) and $I$ (in blue or red). (a) $H>0$ and $I>0$, (b)$H>0$ and $I<0$,(c) $H<0$ and $I>0$, (d) $H<0$ and $I<0$.}\label{fig:5}
\end{center}
\end{figure}

This effect is illustrated in Fig. \ref{fig:5}(a) which shows a sketch of the strip cross-section (in gray), of the electrical current and the associated Oersted field distribution for $I>0$ (in blue), and of the spin-wave modal profile (in full lines and dotted lines for $k>0$ and $k<0$ respectively). Note that the sketch of modal profiles corresponds to the anomalous distribution of the dynamic magnetization across the film thickness described in reference \onlinecite{Kostylev2013}. For simplicity, we consider in this figure a film which is continuous in the plane. The Oersted field of the current flowing along the film is anti-symmetric across the film thickness: it varies linearly across the thickness and has two maxima (a positive and a negative) at the two opposite film surfaces. Due to this contribution, the total static magnetic field inside the film is thickness non-uniform.  From Fig. \ref{fig:5}(a), it is clear that the Oersted field suitably weighted by the spin-wave modal profile does not average out to zero but to a finite value defined as $\delta H_{Oe}$ in the figure. For $k>0$, the resultant field adds to the external field, so that the frequency is increased, whereas for $k<0$ it subtracts from $H$ so that the frequency is decreased. Naturally, the effect reverses when $I$ is reversed (see Fig. \ref{fig:5}(b)). The effect is therefore odd in $I$ and $k$, similar to CISWDS [see Eq. (\ref{eq:1})]. For the mode profiles sketched in Fig. \ref{fig:5}, the sign of this effect is such that it compensates CISWDS (for a positive spin polarization, the Doppler effect shifts the frequency up when the spin wave phase velocity is co-aligned with the electron flow, i.e. when it is anti-aligned to the current). Figures \ref{fig:5}(c,d) illustrate the situation when $\bf{H}$ is reversed. In that case, the modal profile asymmetry is reversed, and the resultant Oersted field reverses accordingly. However, because the static magnetic field points now in the opposite direction, the magnitudes of the total field are the same as in Fig. \ref{fig:5}(a,b). Consequently, the non-reciprocal Oersted field contribution to the spin wave frequencies does not change upon the reversal of $\bf{H}$, similarly to the CISWDS contribution. From this discussion, it is clear that in order for the Oersted-field induced frequency non-reciprocity to appear, the waves should possess asymmetry of modal profiles across the film thickness together with modal-profile non-reciprocity, i.e. the profile asymmetry should be different for the waves propagating in the two opposite directions.

Let us now discuss the origin of this modal-profile non-reciprocity. As a first approach, one can refer to the standard Damon-Eshbach (DE) picture of Magnetostatic Waves. Neglecting the exchange interaction, one obtains a wave with a surface character,\cite{Damon1961} which means that the maximum of the amplitude of magnetization precession is located at one of the film surfaces and this maximum moves to the opposite surface upon switching the direction of wave propagation. The profile of the dynamic magnetization across the film thickness (in the direction $x$) for the DE wave is exponential [$exp(-kx)$]. The decrement is equal to the in-plane wave number $k$. Thus, the wave surface character increases with an increase in the wave number. From this exponential character of the wave profile it follows that for $k>>6\pi/L$ , where $L$ is the film thickness, the wave does not feel the presence of the opposite film surface. Thus, the wave properties should depend entirely on the conditions near the surface at which it is localized. As follows from the DE dispersion law for the ordinary case of the thickness uniform internal field, the frequency of the DE wave increases with an increase in the field magnitude. Therefore, one may expect that in our case of a thickness non-uniform internal field the wave has a larger frequency when it travels along the surface at which the total internal field is maximum and a smaller frequency when it travels along the surface at which the total internal field is minimum. One may also expect that this picture is valid not only for $k>>6\pi/L$ but in the whole $k$-value range and that  the magnitude of this effect increases with an increase in $k$, since the surface character of the wave increases with $k$. These predictions are in full qualitative agreement with the experimental observations. However, a problem arises when one attempts to predict the sign of the Oersted-field induced frequency shift based on this model. Indeed, the modal profile asymmetry of the DE wave is such that the wave-vector $k$ for the wave localized at the surface with internal normal $\mathbf{n}$  verifies:\cite{LeGurevich}
\begin{equation}
\mathbf{k}/|k| =\mathbf{n} \times \mathbf{M}/|\mathbf{M}|, \label{eq:nonrec}
\end{equation}
where $\times$ is the cross-product operation. This relation actually provides a modal profile non-reciprocity opposite to what is sketched in Fig. \ref{fig:5}.\footnote{For example, for $M//e_y$, Eq. (\ref{eq:nonrec}) predicts that the $-k$ wave is localized close to the top surface, whereas it was assumed in Fig. \ref{fig:5}(a) that it is localized close to the bottom surface} Therefore, the OFIFS for a DE wave is expected to have the same sign as the Doppler frequency shift and to increase the total value of the shift. This conclusion is in complete disagreement with our experiment which shows that a concurrent process \textit{compensates} the Doppler frequency shift. This contradiction is removed if the exchange interaction is taken into account. In that case the sign of the modal profile non-reciprocity is determined not only by the sigh of $k$ [as follows from Eq.(\ref{eq:nonrec})], but also by the magnitude of $k$\cite{Kostylev2013} such that in a broad range of wave numbers the spin wave is characterized by the modal profiles shown in  Fig. \ref{fig:5}. In the next chapter, we give a full theoretical calculation of the OFIFS for this situation.

\section{THEORY}\label{sec:theory}

In this chapter we develop a theory to describe OFIFS. Firstly we find that the largely used Eq. (\ref{eq:nonrec}) is not valid in our case since it gives the wrong sign for the OFIFS and the wave should be actually localized at the surface opposite to the one predicted by Eq. (\ref{eq:nonrec}). From this point of view, the sign of the OFIFS contribution found in our experiment represents a strong experimental evidence of the anomalous surface spin wave localization in large-magnetic-moment thin metallic ferromagnetic films described in Ref. \onlinecite{Kostylev2013} and to which the distributions of the dynamic magnetization shown in Fig. \ref{fig:5} correspond.

Then we use the constructed theory to calculate the total frequency shift for our experimental conditions and also to predict the wave number and film thickness ranges where OFIFS contribution to the total frequency shift is not negligible. In the end of the section we briefly discuss the influence of the finite width of the strip on OFIFS and also how the presence of the frequency non-reciprocity of waves originating from the surface magnetic anisotropy of the film (see  discussion in subsection \ref{subsec:psws}) affects the extraction of the current-induced frequency shift.

\subsection{Initial equations}\label{subsec:equations}

To construct the theory we will use the frame of reference shown in figures \ref{fig:1} and \ref{fig:5} (same as in Ref. \onlinecite{Kostylev2013}) and CGS units. The film is assumed to be infinite in the \textit{y}-direction and magnetized to saturation by an external magnetic field $\mathbf{H}=H \mathbf{e}_y$  ($\mathbf{e}_y$ is the unit vector in the \textit{y}-direction). The dynamic magnetization vector $\mathbf{m}$  has only two non-vanishing components ($m_x, m_z$)  which are perpendicular to the static (equilibrium) magnetization vector $\mathbf{M}=M_s (\mathbf{H}/ |H|) \mathbf{e}_y$. The direction $z$ is in the film plane along the spin wave wave vector.  The dynamic effective field $\mathbf{h}_{eff}$ which enters Landau-Lifshitz equation\cite{LeGurevich} has two components: the exchange field $\mathbf{h}_{ex}$ and the dipole field $\mathbf{h}_d$.
For the dynamic magnetization and field in the form of a plane spin wave with a wave number $k$  ($\textbf{k}=k \textbf{e}_z$) and a frequency $\omega$ travelling along $z$ we may use
\begin{equation}
\mathbf{m}, \mathbf{h}_{eff}=\mathbf{m}_k, \mathbf{h}_{effk} exp(i \omega t - i k z).
\end{equation}
We present the dynamic dipole field as a tensor Green's function  $\mathbf{G}_k$ of dynamic magnetization:\cite{Kalinikos1981}
\begin{equation}
\mathbf{h}_{dk}(x)=\int_0^L \mathbf{G}_k(x-x') \mathbf{m}_k(x') dx'  = \mathbf{G}_k \otimes  \mathbf{m}_k,
\end{equation}
where the symbol $\otimes$ denotes the convolution operation and $L$ is the film thickness.  A co-ordinate transformation
\begin{equation}
m_{xk}=(m^{(1)}_k+m^{(2)}_k)/2, \quad m_{yk}=(m^{(1)}_k-m^{(2)}_k)/(2i)
\end{equation}
and a similar transformation for the components of $h_{effk}$
reduce the linearized Landau-Lifshitz equation to a system of integro-differential equations
\begin{eqnarray}
\lefteqn{\omega \mathbf{m}_k =} \label{eq:5} \\
&&  \left|
 \begin{array}{cc} {-[\omega_H+\gamma H_{Oe}-\omega_M(\alpha \partial^2 / \partial x^2 + \alpha k^2+ 1/2) ] \delta}  & {\omega_M(G_q+G_p-\delta/2))} \\
 {\omega_M(G_q-G_p+\delta/2))}  & {[\omega_H+\gamma H_{Oe}-\omega_M(\alpha \partial^2 / \partial x^2+ \alpha k^2+ 1/2) ] \delta}
\end{array}
\right| \otimes \mathbf{m}_k. \nonumber
\end{eqnarray}
In these equations $H_{Oe}=J_0 J(x-L/2)$, where $J$ is measured in $A.cm^{-2}$ and $J_0=4 \pi/10$ is a factor converting the Oersted field into Gaussian units, $\alpha$ is the exchange constant, $\delta=\delta(s)$ is the Dirac delta function, $\omega_H=\gamma H$, $\omega _M= (H/|H|)\gamma 4\pi M_s$, and $\gamma$ is the gyromagnetic coefficient for the magnetic material.  The column vector $\mathbf{m}_k$ has now components $(m_k^{(1)},m_k^{(2)})$.
In the form which is the most convenient for the analysis below, the components of the Green's function are presented in Ref. \onlinecite{Kostylev2013}. They are as follows:
\begin{eqnarray}
G_p(s)=\frac{\mid k \mid}{2} exp(-\mid k \mid |s|) \label{eq:6}  \\
G_q(s)=sign(s) \frac{k}{2} exp(-\mid k \mid |s|),\label{eq:7}
\end{eqnarray}
where $s=x-x'$.
One sees that the eigenfrequency of spin waves $\omega$  in Eq. (\ref{eq:5}) represents an eigenvalue of the integro-differential operator. The terms involving the exchange constant $\alpha$ originate from the exchange contribution to the spin wave frequency (see Refs \onlinecite{Kalinikos1981} and \onlinecite{Kostylev2013} for details.)

\subsection{Solution}\label{subsec:solution}

A valid way to solve Eq. (\ref{eq:5}) is by treating the terms $\gamma H_{Oe}$ as a small perturbation of the integro-differential operator for $H_{Oe}=0$ ("unperturbed operator"). Then one can use the well established theory of perturbation of eigen-values to obtain OFIFS. The calculation is especially simple in the exchange-free case $\alpha=0$. We perform it in the appendix. This calculation clearly demonstrates that the origin of OFIFS is the combination of  the surface character of the exchange-free DE wave and of the modal profile non-reciprocity.
As shown in Ref. \onlinecite{Kostylev2013}, the Damon-Eschbach theory is valid for description of the \textit{modal profiles} of the spin waves in ferromagnetic films, provided the films are relatively thick and the wave number is large, such that the frequency of the first exchange standing  spin wave mode (1st SSW) falls within the frequency range of the existence of the exchange-free DE wave and the frequency for that particular wave number lies above the frequency for the 1st SSW.

The 1st SSW branch enters the spectrum of the DE wave for the "critical" thickness
\begin{equation}\label{eq:8}
L^2_c=\alpha \pi^2/ (\sqrt{\nu^2+\nu+0.5}-\nu-0.5),
\end{equation}
where $\nu=H/(4\pi M_s)$.
In our experiment $\nu$=0.028 and $\alpha=3.1 \times 10^{-13}$ which gives  $L_c=39$ nm. Thus, for the samples with thicknesses $L>39$ nm one may expect a range of wave numbers where Eq. (\ref{eq:30}) in the appendix is valid and OFIFS is negative. This wave number range increases with an increase in the thickness. For instance, for $ L=70$ nm and all other  parameters as in our experiment the 1st SSW branch intersects the DE branch at $k=2$ $\mu m^{-1}$. Thus, for the most of the wave number range accessible with the travelling wave spectroscopy OFIFS will be negative and in agreement with Eq. (\ref{eq:30}) for a film this thick.

The thickness of the thickest sample in our experiment is 40 nm which is quite close to $L_c$. Therefore, for this particular sample as well as for all films from the most technologically important thickness range 40 nm and below, one has to include the exchange interaction in the theory. We use the same initial equations from subsection \ref{subsec:equations} to construct the theory. The theory is based on solving these equations using the Boubnov-Galerkin method. This method consists in the expansion of $\bf{m}_k$ in a Fourier series.\cite{Kalinikos1981} For simplicity, we assume the "unpinned surface spins" exchange boundary conditions on both film surfaces. In this case the system of cosine functions is the natural choice of a full ortho-normal basis of functions which satisfy the boundary conditions. We also assume that the frequency of the 2nd SSW branch is well above the upper frequency limit for existence of the exchange-free DE wave. Therefore, only the first two terms of the  series expansion should be taken into account (see Eq. (47) in Ref. \onlinecite{Kalinikos1986} for the explanation):
\begin{equation}
\textbf{m}_k(x)=\textbf{m}_{k0}+\sqrt{2}\textbf{m}_{k1} cos(\pi x /L).
\end{equation}
We substitute this solution into Eq. (\ref{eq:5}) and project the resultant equation on the ortho-normal basis of these cosine functions. As a result we obtain a system of four algebraic equations.  The matrix $C_k$ of the coefficients of this system of equations has the form as follows:
\begin{equation}\label{eq:10}
C_{k} =  \left|
 \begin{array}{cccc}
{ -A_{00} - \omega}  & {-B_{00}}  & {-A_{01}} & { B_{01} } \\
{B_{00} }  & {A_{00} -\omega}  & { B_{01}} &  { A_{01} } \\
{ -A_{01}} & { -B_{01}} & {-A_{11} - \omega}  &{ -B_{11}  } \\
{-B_{01} }& { A_{01}} & { B_{11}}  & { A_{11} -\omega  } \\
\end{array}
\right| ,
\end{equation}
where $A_{00}=\omega_H+\omega_M\alpha k^2+\omega_ M/2$, $B_{00}=[1/2-P_{00}]\omega_M$, $A_{11}=A_{00}+\omega_M\alpha (\pi /L)^2$, $B_{11}=[1/2-P_{11}]\omega_M$, $B_{01}=\omega_MQ_{01}$, $A_{01}=\gamma J J_0 d_{01}$.

The quantities $P_{00}$, $P_{11}$ and $Q_{01}$ are particular cases of the dipole elements $P_{nn'}$ and $Q_{nn'}$ derived in Ref. \onlinecite{Kalinikos1981}. They are obtained by projecting $G_p$  and $G_q$ [Eqs. (\ref{eq:6}) and (\ref{eq:7}) respectively] on the basis of the cosine functions. These quantities have the forms as follows
\begin{eqnarray}
P_{00}=1-[1-exp(-|k|L)]/(|k|L), \\
P_{11}=(kL)^2[1-2(|k|L) \frac{1+exp(-|k|L)}{(kL)^2+\pi^2}], \\
Q_{01}=-2 \sqrt{2} (kL) \frac{1+exp(-|k|L)}{(kL)^2+\pi^2}.
\end{eqnarray}

The quantity $d_{01}$ is obtained in a similar way by projecting the thickness dependence of $H_{Oe}$ onto the same basis. The Oersted-field element reads:
\begin{equation}
d_{01}=-2 \sqrt{2}  L /\pi^2.
\end{equation}
The Oersted field is anti-symmetric across the film thickness, similar to $G_q$. Therefore only the (0,1)  and (1,0) components of $d$ are not vanishing, similar to the $Q$-elements.

The eigenfrequencies of spin waves are given by the condition $det(C_k)=0$. Evaluating this determinant analytically we obtain a dispersion relation in the presence of the DC current. This relation can be cast in the following form:
\begin{equation}\label{eq:15}
(\omega_1^2-\omega^2)(\omega_0^2-\omega^2)+4 \gamma J J_0 \omega \omega_M^2 d_{01} Q_{01} (P_{00}-P_{11}) =0.
\end{equation}
In this equation $\omega_0$  and $\omega_1$ are the (positive) roots of the bi-quadratic equation which represents the dispersion relation for $J=0$:
\begin{eqnarray}\label{eq:16}
(\omega_{11}^2-\omega^2)(\omega_{00}^2-\omega^2) \nonumber \\
-\omega_M^2Q_{01}^2[(\omega_{11}^2-\omega^2)+(\omega_{00}^2-\omega^2)]+ \omega_M^4[Q_{01}^4+Q_{01}^2(P_{00}-P_{11})^2 -\alpha^2 (\pi/L)^4]=0,
\end{eqnarray}
and $\omega_{00}$ and $\omega_{11}$ are the positive roots of the determinants of the upper and lower 2x2 diagonal  blocks  of the block matrix (\ref{eq:10}) respectively. These roots are given by the expressions as follows
\begin{eqnarray}
\omega_{00}^2=[\omega_H+\omega_M (\alpha k^2+P_{00})][\omega_H+\omega_M (1+\alpha k^2-P_{00})], \\
\omega_{11}^2=[\omega_H+\omega_M (\alpha( k^2+(\pi/L)^2)+P_{11})][\omega_H+\omega_M (1+\alpha( k^2+(\pi/L)^2)-P_{11}]).
\end{eqnarray}
Note that in order to obtain the dispersion relation in the simple form (\ref{eq:15}) we neglected the terms of the second order in $J$ because $H_{Oe}^2<<(4\pi M_s)^2$.

The frequency shift due to the presence of the DC current is small compared to the unperturbed spin-wave frequency (\ref{eq:16}). Therefore we may assume that $\omega=\omega_0+\delta \omega$, where $|\delta \omega|<< \omega_0$. This allows one to expand Eq. (\ref{eq:15}) in Taylor series in $\delta \omega$. Keeping only the linear terms of this expansion we obtain a very simple  formula for OFIFS:\footnote{This expression of the OFIFS is for CGS units ($L$ should be taken in cm and J in $A/cm^2$). The same formula can be used in SI units provided $J_0$ is replaced by $\mu_0$.}
\begin{equation}\label{eq:19}
\delta \omega = \frac{2 \gamma J J_0 \omega_M^2 d_{01} Q_{01} (P_{00}-P_{11})}{\omega_1^2-\omega_0^2}.
\end{equation}

\begin{figure}[!htp]
\begin{center}
\includegraphics[scale=0.70]{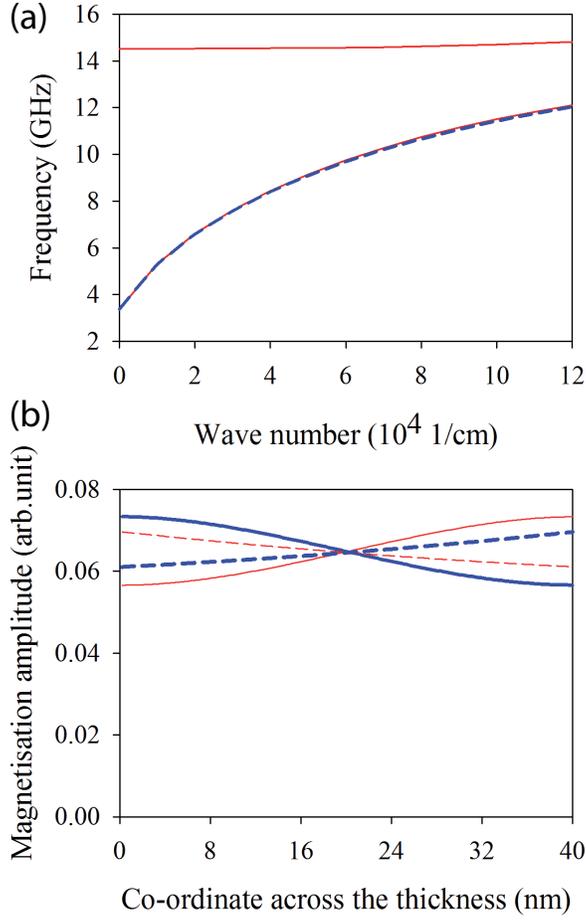}
\caption{(a) Dispersion of dipole-exchange spin waves in a 40nm-thick permalloy film for $I=0$. Parameters of calculation. Film thickness: 40 nm, saturation magnetization: $4\pi M$=10500 G, gyromagnetic coefficient: $\gamma/ 2 \pi$=2.8 MHz/Oe, internal static magnetic field is 137.2 Oe, exchange constant $A=1.355 \times 10^6$ erg/cm ($\alpha=3.04 \times 10^{-13} cm^2$).  Dashed line: exchange-free Damon-Eshbach dispersion law (given here for comparison). (b) Modal profiles of the fundamental mode of the dipole-exchange waves (solid lines) and of exchange-free Damon-Eshbach waves (dashed lines). Thick lines: $k>0$, thin lines: $k<0$.}\label{fig:6}
\end{center}
\end{figure}

Let us analyze this expression. $P_{00}$ and $P_{11}$ are positive, $P_{00}>P_{11}$, and $d_{01}$ is negative. The sign of $Q_{01}$ changes upon switching the direction of the wave vector. $Q_{01}$ is negative for $k>0$. The unperturbed dispersion for the case of our sample ($L=40$ nm) given by Eq. (\ref{eq:16}) is shown in Fig. \ref{fig:6}(a). One sees that for this thickness the 1st SSW lies within the frequency band of existence of the DE wave, in agreement with Eq. (\ref{eq:8}). Hybridization of the DE wave and of the 1st SSW results in repulsion of the branches. As one sees from this graph, all the wave numbers for which the experimental data were taken correspond to the lower branch of this spectrum. To calculate OFIFS for this branch one has to assume that $\omega_0$ is the frequency which corresponds to it and $\omega_1$ is the frequency for the upper frequency branch for the same wave number. This assumption implies that $\omega_0<\omega_1$ and hence $\delta \omega > 0$ for $k>0$. This result is in agreement with our experiment [i.e. the OFIFS is of the opposite sign compared to the Doppler shift for $P>0$, see Eq. (\ref{eq:1})]. On the other hand, the sign is opposite to the result of the exchange-free theory in the appendix.

If we now assume that $\omega_0$  belongs to the upper branch, $\omega_1$ is then the respective frequency from the lower branch, and hence $\delta \omega < 0$. This result is in agreement with the exchange-free theory, as predicted above.

One also notices that the frequency shift scales as $Q_{01}$. As seen from Eqs. (\ref{eq:10}) and (\ref{eq:16}) $Q_{01}$ is responsible for the hybridization and repulsion of the DE and 1st SSW branches. It is also responsible for the surface character of the waves as well as for the modal profile non-reciprocity (all in the absence of the current).\cite{Kostylev2013} Indeed, the asymmetry parameter $s$ for the modal profile\cite{Kostylev2013} scales as $\omega_M Q_{01}/(\omega_{00}-\omega_{11})$.  Given that $\omega_{00}$ is close to $\omega_0$  and $\omega_{11}$ to $\omega_1$, this term is very close to the factor $\omega_M Q_{01}/(\omega_{0}-\omega_{1})$ which enters Eq. (\ref{eq:19}). This demonstrates that the origin of OFIFS is the  modal profile non-reciprocity, in agreement with the naive picture in Fig. \ref{fig:5} and the exchange-free theory in the appendix. From the comparison of the equation (15) from Ref. \onlinecite{Kostylev2013} for the modal profile asymmetry and Eq. (\ref{eq:19}) one finds that the positive $\delta \omega$ corresponds to the anomalous wave localization, i.e. localization of the wave at the surface opposite to one at which the exchange-free DE wave is localized. Similarly, the negative $\delta \omega$ corresponds to the normal localization [Eq. (\ref{eq:nonrec})]. The calculated profiles for the dipole-exchange waves are shown in  Fig. \ref{fig:6}(b). For comparison, the respective modal profiles calculated with the exchange-free theory are also displayed in this figure. Note that the sketches of the profiles in Fig. \ref{fig:5} are for the dipole-exchange waves.

The dependence of OFIFS on the wave number is quite steep. If one expands the product $Q_{01}(P_{00}-P_{11})$ in the numerator of Eq. (\ref{eq:19}) into Taylor series in $k$, one obtains that for $kL<<1$ the leading term of the expansion is the $(kL)^2$-term. The leading term of the Taylor expansion of the denominator is $k^0$-one. Thus, the dependence of OFIFS is at least parabolic.

\subsection{Numerical results}\label{subsec:numerical}

\begin{figure}[!htp]
\begin{center}
\includegraphics[scale=0.70]{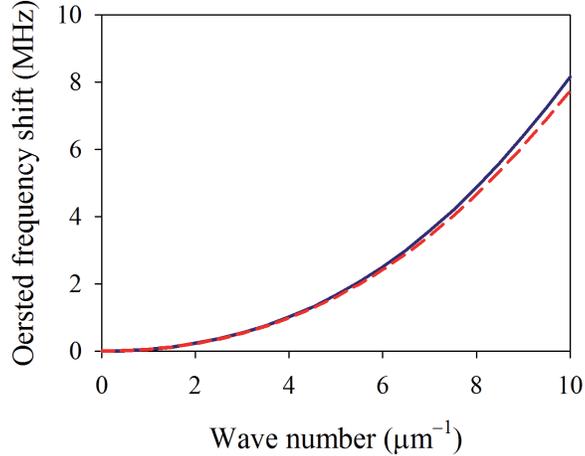}
\caption{Calculated Oersted-field induced frequency shift for dipole exchange spin waves for a current density $J=6.67$~$10^6$ $A/cm^2$. Solid line: Internal field is 137.2 Oe. Dashed line: internal field is 280 Oe. Other parameters of calculation are the same as in Fig. \ref{fig:6}. Eq. (\ref{eq:19}) was used to produce these data.}
\label{fig:7}
\end{center}
\end{figure}

The result of our calculation by using Eq. (\ref{eq:19}) is shown in Fig. \ref{fig:7} for a current density of 6.67 $10^6$ $A/cm^2$. We perform this calculation for the parameters of the film we use in our experiment (solid line, see the figure caption for the details). The Oersted field induced frequency shift grows quite quickly with an increase in the wave number. Given the scaling law above [$(kL)^2$ for very small $k$  and steeper for larger $k$-values], this suggests that for thick films OFIFS may become dominating for large wave numbers, since the Doppler frequency shift $\Delta f_{Dop}$ scales linearly with $k$ and is independent from $L$. In the same figure we also show the result of the calculation for a larger internal field (280 Oe, dashed line). One sees that the applied field does not have a significant effect on the magnitude of OFIFS.

The curves in Fig. \ref{fig:4} are the calculated \textit{total} frequency shift $\Delta f_{tot}=\Delta f_{Dop}+\delta f_{Oe}$ for for the 40 nm and 10 nm-thick films assuming a degree of spin polarization of the current $P=0.6$. The other parameters of calculation are the same as for Fig. \ref{fig:6}. Here $\delta f_{Oe}$ is deduced from Eq. (\ref{eq:19}) as $\delta f_{Oe}=-\delta \omega/(2 \pi)$, to comply with the sign convention used in the definition of $\delta f_{odd}$. From this figure one sees that the dependence deviates from the linear  one with a negative slope starting from very small wave numbers and the total shift becomes negative for $k>6$ $\mu m^{-1}$. Starting from this $k$-value OFIFS represents the dominating contribution to the total frequency shift.

\begin{figure}[!htp]
\begin{center}
\includegraphics[scale=0.70]{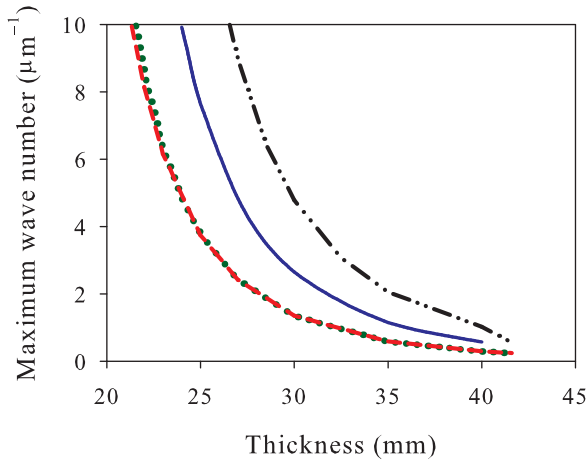}
\caption{Characteristic value of the spin wave number for which the Oersted field induced frequency shift starts to give contribution to the total frequency shift of particular magnitude. Solid line: $\delta f_{Oe}/ \Delta f_{Dop}=10 \%$, internal field is 137.2 Oe; dashed line: $\delta f_{Oe}/ \Delta f_{Dop}=5 \%$, internal field is 137.2 Oe; dash-dot-dotted line: $\delta f_{Oe}/ \Delta f_{Dop}=10 \%$, internal field is 5 kOe. Dotted line: approach from Ref. \onlinecite{Sekiguchi2012} of using the difference of group velocities instead of the frequency difference to extract the degree of spin polarization. Internal field is 137.2 Oe and the assumed contribution of the Oersted field induced frequency shift to the total frequency shift is 10 percent. (Note that occasionally this line almost overlaps with the dashed line.)
The Doppler shift is calculated for P=0.6. All  other parameters of calculation are the same as for Fig. \ref{fig:6}. Eq.(\ref{eq:19}) was used to produce these data.}\label{fig:8}
\end{center}
\end{figure}

We also calculate the characteristic value of $k$ for which OFIFS starts to provide a contribution to the total frequency shift of a specific magnitude.  We consider two cases: when $\delta f_{Oe}$ becomes either 5 or 10 percent of the Doppler shift. These data are shown in Fig. \ref{fig:8}. From this figure one sees that the maximum $k$-value, for which the contribution of OFIFS to the measured degree of spin polarization can be regarded as negligible to experimental accuracy, drops very quickly with the film thickness. This is due to the above discussed steep dependence of $\delta \omega$ on $L$. This characteristic wave number also depends on the applied field through the dependence of the mode frequencies in the denominator of Eq. (\ref{eq:19}) on the applied field. However, the dependence is not very strong which is seen from the comparison of  two plots in this figure: for $H=$137.2 Oe and 5 kOe.

In this graph we also compare two competing methods of extracting the spin polarization from the Doppler shift data: ours which is based on the measurement of the frequency and the one from Ref. \onlinecite{Sekiguchi2012} which is based on measurement of the spin wave group velocity. One sees that results of the measurements of the variation in the group velocity due to the presence of a DC current should be much stronger affected by OFIFS than the measurements of the frequency for the same value of $k$. Indeed, the characteristic $k$ for the group velocity measurements allowing a 10-percent contribution of OFIFS to the total frequency shift coincides with the  characteristic $k$-value for the 5-percent contribution of OFIFS  in the frequency measurements.

In this section we also consider another effect which has not been taken into account yet in our theory. This is the non-reciprocity of spin wave dispersion seen for the 40nm-thick sample for $I$=0 (see Fig. \ref{fig:2}). This type of non-reciprocity is usually attributed to the non-uniformity of material parameters across the film thickness.\cite{Amiri2007} One of these potential non-uniformities is surface magnetization pinning which originates from the presence of surface anisotropy. Another possible reason which is worth mentioning in this context is spontaneous formation of a thin surface sublayer whose magnetic properties are different from the bulk of the material (see e.g. discussion in Ref. \onlinecite{Kostylev2010}). Here for simplicity we assume the presence of magnetization pinning at one of the film surfaces. To understand the effect of the single-side pinning we perform direct numerical solution of Eq. (\ref{eq:5}). To obtain the effect of the non-reciprocity, we assume that spins at the film surface facing the antennae are partly pinned and are completely unpinned on the other film surface. This simulation gives the correct values for the spin wave frequencies. For instance, one obtains the same value as in the experiment (10.709 GHz) for  $k_z=+7.8$ $\mu$m$^{-1}$ and $I$=10 mA. For the wave propagating in the opposite direction both theory and experiment give 10.749 GHz. The value of the surface pinning parameter used in this calculation is $1.8 \times 10^5$ cm$^{-1}$. This value corresponds to the value of the constant of the surface normal uniaxial anisotropy of 0.245 mJ/m$^2$.

The solid line in Fig. \ref{fig:4}(b) is actually the total current-induced frequency shift extracted from this numerical calculation, taking into account a partial pinning of the magnetization at one film surface. As in the experiment we calculate and show $\delta f_{odd}$ in Fig. \ref{fig:4}, in order to remove the even contribution to the total frequency shift from the raw simulation data (as discussed in subsection \ref{subsec:CIFS}, an even contribution originates from the interplay between the Oersted field and an asymmetric surface pinning). The dashed line in Fig. \ref{fig:4}(b) is the result of our analytical solution [Eq. (\ref{eq:19})]. This solution assumes unpinned surface spins at both film surfaces and, consequently, no frequency non-reciprocity in the absence of the DC current. The very good agreement of the dashed line with the solid one confirms the validity of our experimental approach for removing the $I=0$-non-reciprocity from the experimental data by calculating $\delta f_{odd}$. The dots in the figure are the available five experimental points. One sees good quantitative agreement with the experiment.

The last point which we want to comment on in this section is the effect of the finite strip width. Our numerical solutions of a 2D version of Eq. (\ref{eq:5}) shows that the presence of the geometrical confinement in the plane of the film does not change the OFIFS qualitatively. The dominating effect of the confinement is a frequency shift for $I=0$ due to the static demagnetizing field which appears because the strip is magnetized along a hard axis. This effect is easily taken into account in our 1D model above of an "effective" continuous film by subtracting some effective demagnetizing field $H_{dem}$ from the applied field in Eq. (\ref{eq:15}). In the example of Fig. \ref{fig:2} we subtract 142.8 Oe from the applied field in order to obtain the good agreement with the experiment. This value is quite close to the value of the static demagnetizing field averaged across the area of the stripe cross-section which we obtain with LLG Micromagnetic Simulator\cite{LLG} for the applied field of 280 Oe in this geometry.

\section{CONCLUSION}\label{sec:conclusion}

In this paper we have studied the current-induced frequency shift for spin waves propagating perpendicular to the direction of the applied field in an in-plane magnetized 40nm-thick Permalloy strip. Contrary to the previous measurements of current-induced spin-wave Doppler shift in thinner films, this experiment revealed a non-monotonic dependence of the extracted degree of spin polarization on the spin wave number. For large wave numbers, the extracted value of the degree of spin polarization is negative, which is unphysical. We suggest that this phenomenon originates from a contribution from a concurrent effect, namely a spin wave frequency non-reciprocity induced by the Oersted field generated by the DC current applied to the sample in order to observe the Doppler effect. This contribution to the total frequency shift is experimentally indistinguishable from the Doppler frequency shift and grows with an increase in the sample thickness.

To confirm this idea, a theory of the Oersted-field induced non-reciprocal frequency shift has been constructed. The theory unambiguously demonstrates the dominating role of this type of frequency non-reciprocity in the formation of the total frequency shift in the presence of a DC current for Permalloy films with thicknesses above 20nm. The comparison of the theory with the experiment also confirms the recent theoretical prediction of the anomalous modal-profile non-reciprocity for large-magnetic-moment metallic ferromagnetic films.\cite{Kostylev2013}

This work allows one to understand the limitations of the technique of the current-induced spin-wave Doppler shift when carried out in the Magnetostatic Surface Wave (or Damon-Eshbach) geometry. We found that this configuration is fully appropriate for film thickness 20nm or below. For thicker films care should be taken in order to avoid the situation where the effect of the Oersted field potentially becomes dominant. As follows from our theory, using small wave vectors is the way to avoid it.

The present case of the 40nm-thick film and large $k$-values is a clear example of such unfavorable experimental conditions. Even in this situation, the current-induced modification of spin-wave propagation can be measured very precisely and interpreted with an explicit analytical theory. In our opinion, this possibility is a natural advantage of using spin waves for probing the spin-transfer torque. Indeed, due to the simple plane-wave structure and linearity of small-signal spin waves, accurate explicit analytical models can be constructed in 2D (and simple numerical algorithms can be developed in 3D).  This is in strong contrast to the more widely studied case of the current-induced domain-wall motion. Since the domain walls are intrinsically nonlinear objects, full (nonlinear) micromagnetic models are  required, even in the simplest cases. For instance, full 3D micromagnetic simulations are necessary to treat the influence of the same effect of the Oersted field on the domain wall dynamics.

\begin{acknowledgments}
This work was supported by the ANR (NanoSWITI, ANR-11-BS10-003) and the Australian Research Council.
\end{acknowledgments}

\appendix

\section{EXCHANGE-FREE THEORY OF THE OERSTED-FIELD INDUCED NON-RECIPROCAL FREQUENCY SHIFT FOR MAGNETOSTATIC SURFACE WAVES}\label{appendix}

The exchange-free theory for the magnetostatic surface waves was first suggested by Damon and Eshbach\cite{Damon1961} more than 50 years ago.  The straightforward way to obtain this result is by solving the second-order partial derivative equation - called Walker Equation-\cite{LeGurevich} employing the appropriate electrodynamic boundary conditions.  (Walker equation is derived by solving  Landau-Lifshitz Equation for the magnetic torque together with Maxwell equations in the magnetostatic approximation.)  Although the Walker-equation approach is the standard way to tackle the spin wave dispersion problem, in this paper we will follow a different route: we will solve the eigenvalue problem for the same system of  integral equations (\ref{eq:5})-(\ref{eq:7}). An exact analytical solution exists for this system for the vanishing DC current $\mathbf{I}$ and $\alpha=0$. We will employ this solution as the zero approximation to construct the perturbation theory for non-vanishing values of $\bf{I}$. The Green's function approach makes the perturbation theory especially simple.

The analytical solution of the system (\ref{eq:5})-(\ref{eq:7}) for $\alpha=0$ and $\textbf{I}=0$  has the form
\begin{equation}
\mathbf{m}_k=\left| \begin{array} {c} m^{(1)}_k \\
 m^{(2)}_k \end{array} \right| =
\left| \begin{array} {c} B exp(-k x) \\
- A exp(k x-L) \end{array} \right|
\end{equation}
After substitution of this solution in Eq. (\ref{eq:5}) and some straightforward algebra one finds that the two eigenvalues $\omega_1=+\omega_0$ and $\omega_2=-\omega_0$ ($\omega_0>0$) of the system of the integral equations (\ref{eq:5})-(\ref{eq:7}) are given by the Damon-Eschbach dispersion relation\cite{Damon1961}
\begin{equation}\label{eq:21}
\omega_0^2=\omega_H (\omega_H+\omega_M)+\frac{\omega_M^2}{4} (1-exp(-2 k L)).
\end{equation}
The respective \textit{right-hand} (column) eigen-vectors are as follows:
\begin{equation}\label{eq:22}
|\mathbf{m}_1>=
 \left| \begin{array} {c} exp(-k x) \\
 -\frac{2 \omega_{H}-2\omega_0+\omega_{M}}{\omega_{M}} exp(k x)  \end{array} \right|,
 \end{equation}
\begin{equation}\label{eq:23}
|\mathbf{m}_2>=
 \left| \begin{array} {c} exp(-k x) \\
 -\frac{2 \omega_{H}+2\omega_0+ \omega_{M}}{\omega_{M}} exp(k x)  \end{array} \right|.
\end{equation}

From Eq. (2) it follows that for a given $k>0$ the eigen-wave with the positive frequency $\omega_1$ propagates in the positive direction of the axis $z$ and the wave with the negative eigen-frequency $\omega_2$ travels in the opposite direction. In the following we will use this convention (a positive or negative $\omega$ for an always positive $k$) to identify the waves travelling in particular directions along the axis $z$. This way  is natural, given the role of $\omega$ as an eigenvalue of Eq. (\ref{eq:5}).

The surface character of the waves follows from Eqs. (\ref{eq:22}-\ref{eq:23}). From Eq. (\ref{eq:21}) it follows that $\omega_0$ never exceeds the limiting value $\omega_H+\omega_M/2$. Therefore one finds that $\frac{2 \omega_{H}+2\omega_0+\omega_{M}}{\omega_{M}} exp(k x)$ is always larger (and usually significantly larger) than $exp(-kx)$, and that $\frac{2 \omega_{H}-2\omega_0+\omega_{M}}{\omega_{M}} exp(k x)$ is (significantly) smaller than $exp(-kx)$. Thus, the wave with the positive frequency is localized near the lower film surface $x=0$  [Eq. (\ref{eq:22})] and the wave with the negative eigen-frequency is localized at the upper film surface $x=L$ [Eq. (\ref{eq:23})].

For the further analysis we will also need the respective left-hand eigen-vectors. (Since the integral operator of this equation is not symmetric, the right-hand eigen-vectors are not orthogonal to each other but are orthogonal to the respective left-hand eigen-vectors.) The  left-hand (row) eigen-vectors are given by the following equations:
\begin{equation}\label{eq:24}
<\mathbf{m}_1|=
 \left| \begin{array} {ccc} \frac{2 \omega_{H}+2\omega_0+\omega_{M}}{\omega_{M}}  exp[k (L-x)] & ; &
 exp[k (x-L)] \end{array} \right|,
\end{equation}
\begin{equation}\label{eq:25}
<\mathbf{m}_2|=
 \left| \begin{array} {ccc} \frac{2 \omega_{H}-2\omega_0+\omega_{M}}{\omega_{M}}  exp[k (L-x)]  & ; &
 exp[k (x-L)]  \end{array} \right|.
 \end{equation}
One can easily see that these sets of vector functions are orthogonal: $\int^L_0 <\mathbf{m}_i(x)|\mathbf{m}_j(x)>dx=0$ for $ i \ne j$.

Now we introduce the Oersted field $H_{Oe}$ of a DC current $\mathbf{I}$ with a current density $\mathbf{J}$ flowing along the axis $z$. This field has only one component $H_{Oey}=H_{Oe}=J_0 J(x-L/2)$ which is anti-symmetric with respect to the half-thickness of the film [$x=L/2$, see Fig. (\ref{fig:5})].   This field combines with the applied field $\mathbf{H}$. This leads to modification of the term $\omega_H$ in Eq. (\ref{eq:5}). This term now reads:
\begin{equation}
\omega_H=\gamma(H+H_{Oe}).
\end{equation}
Since $H>>H_{Oe}$ we may treat $\gamma H_{Oe}$ as a perturbation term for the original Eq. (\ref{eq:5}) (i.e. for the equation with $\omega_H=\gamma H$). This operator perturbation gives rise to a perturbation of eigen-values of the original operator, i.e. to an eigen-frequency shift $\delta \omega(I)=\omega(I)-\omega(I=0)$. In the first approximation this frequency shift reads:
\begin{equation}\label{eq:27}
\delta \omega_i=\int^L_0 <\mathbf{m}_i(x)|\delta C(x)|\mathbf{m}_i(x)>dx \quad / \quad \int^L_0 <\mathbf{m}_i(x)|\mathbf{m}_i(x)>dx,
\end{equation}
where $i=1,2$ indexes the unperturbed eigen-values  and eigen-functions (see Eqs. (\ref{eq:21}-\ref{eq:23})) and  $\delta C(x)$ is the operator of the perturbation:
\begin{equation}
\delta C(x) =\gamma J J_0 (x-L/2)  \left|
 \begin{array}{cc} -1  &  0 \\
0  &  1
\end{array}
\right|.
\end{equation}
 Calculation of the integrals in  Eq. (\ref{eq:27}) reveals that $\delta \omega$ is even in frequency:
\begin{equation}
\delta \omega_2(I)=\delta \omega_1(I)= \delta \omega(I)= \gamma J J_0 [1 - k L \quad coth(kL)]/(2k)
\end{equation}
 As a result,  the total frequency shift due to the Oersted field reads:
\begin{equation}\label{eq:30}
\delta \omega_{Oe}=\frac{\omega_1+\delta \omega(I) - \mid \omega_2+\delta \omega(I) \mid}{2} = \delta \omega(I)=  \gamma J J_0 [1 - k L \quad coth(k L)]/(2k).
\end{equation}

Several important conclusions can be drawn from this expression. Firstly, Eq. (\ref{eq:30}) demonstrates the important role of the wave profile nonreciprocity in the formation of the Oersted frequency shift. Indeed, the magnitude of the shift is given by the projection of the thickness-profile of the perturbation (which is anti-symmetric in $x-L/2$) on the basis of the modal profiles of the eigen-waves given by Eqs. (\ref{eq:22})-(\ref{eq:23}). These profiles are completely uniform (symmetric) across the film thickness for $k=0$. As a result, $\delta \omega_{Oe}=0$ for $k=0$. With an increase in $k$  the frequency shift scales as
\begin{equation}\label {eq:31}
\delta \omega_{Oe} \simeq -\gamma J J_0  L^2 k [1 -(k L)^2/15 ]/6.
 \end{equation}
Thus, its magnitude grows with an increase in $k$ in a way similar to the increase in the surface character of the waves with the increase in $k$. One also sees that the effect is odd in $I$ and does not depend on the applied field, hence it is even in $H$. The parity properties of the expression Eq. (\ref{eq:31}) are in full agreement with what one observes in the experiment on the 40nm-thick sample. The magnitude of the current-induced shift is comparable to the Doppler shift. What is in the complete disagreement with the experimental data is the sign of the effect. In Eq. (\ref{eq:31}) it is the same as for the Doppler frequency shift. However, the experimental $k$ dependence of the total frequency shift for the 40 nm-thick sample demonstrates that the Doppler shift and the Oersted-field induced one should be of the opposite signs, in order for $\delta f_{Oe}$ to overcompensate $\Delta f_{Dop}$ for larger $k$ values and thus to change the sign of the total frequency shift between $k$=3.9 and 7.8 $\mu$m$^{-1}$.

As shown in Ref. \onlinecite{Kostylev2013}, the exchange-free theory of the modal-profile non-reciprocity is valid for the sections of the dispersion curve for the Damon-Eschbach waves which are located \textit{above} the frequency of the first exchange standing spin wave mode (1st SW). This mode enters the frequency range of existence of the Damon-Eshbach branch for thick Permalloy films and intersects the Damon-Eshbach dispersion at large wave numbers. This implies that the exchange-free theory of OFIFS is valid for thick films and large wave numbers. For smaller wave numbers and thinner films one always has to use the theory which includes the exchange interaction, as provided in Section \ref{sec:theory}.


\end{document}